\documentclass[aps,prb,twocolumn,superscriptaddress,showpacs,preprintnumbers]{revtex4-1}
\usepackage{graphicx }% Include figure files
\usepackage{dcolumn}% Align table columns on decimal point
\usepackage{bm}% bold math
\usepackage{braket}
\usepackage{amsmath}
\usepackage[caption=false]{subfig} 
\usepackage{setspace} 
\usepackage{tabularx}
\usepackage{color}

\newcolumntype{Y}{>{\centering\arraybackslash}X}

\begin{document}

%\preprint{insert preprint}

\title{Microscopic coexistence of superconductivity and magnetism in Ca$_{1-x}$Na$_x$Fe$_2$As$_2$}
%: A system with ferromagnetic quantum criticality}% Force line breaks with \\
\author{Philipp Materne}
\author{Sirko Kamusella}
\author{Rajib Sarkar}
\author{Til Goltz}
\author{Johannes Spehling}
\author{Hemke Maeter}
\affiliation{Institute of Solid State Physics, TU Dresden, D-01069 Dresden, Germany}
\author{Luminita Harnagea}
\affiliation{% Leibniz-Institut für Festkörper- und Werkstoffforschung Dresden, 01069, , Germany\\
Leibniz Institute for Solid State and Materials Research (IFW) Dresden, D-01069, , Germany}
\author{Sabine Wurmehl}
\author{Bernd B\"uchner}
\affiliation{Institute of Solid State Physics, TU Dresden, D-01069 Dresden, Germany}
\affiliation{% Leibniz-Institut für Festkörper- und Werkstoffforschung Dresden, 01069, , Germany\\
Leibniz Institute for Solid State and Materials Research (IFW) Dresden, D-01069, , Germany}
\author{Hubertus Luetkens}
\affiliation{Laboratory for Muon Spin Spectroscopy, Paul Scherrer Institute, CH-5232 Villigen, Switzerland}
\author{Carsten Timm}
\affiliation{Institute of Theoretical Physics, TU Dresden, D-01069 Dresden, Germany}
\author{Hans-Henning Klauss}
\affiliation{Institute of Solid State Physics, TU Dresden, D-01069 Dresden, Germany}
\altaffiliation[]{h.klauss@physik.tu-dresden.de}%Lines break automatically or can be forced with \\

%\affiliation{}

\date{\today}% It is always \today, today,

\begin{abstract}
We present a detailed investigation of the magnetic and superconducting properties of Ca$_{1-x}$Na$_x$Fe$_2$As$_2$ single crystals with \textit{x}~=~0.00, 0.35, 0.50, and 0.67 by means of the local probe techniques M\"ossbauer spectroscopy and muon spin relaxation experiments.
With increasing Na-substitution level, the magnetic order parameter as well as the magneto-structural phase transition are suppressed.
For \textit{x}~=~0.50 we find a microscopic coexistence of magnetic and superconducting phases accompanied by a reduction of the magnetic order parameter below the superconducting transition temperature $T_c$.
A systematic comparison with other 122 pnictides reveals a square-root correlation between the reduction of the magnetic order parameter and the ratio of the transition temperatures, $T_c/T_N$, which can be understood in the framework of a Landau theory.
In the optimally doped sample with $T_c~\approx~34$~K, diluted magnetism is found and the temperature dependence of the penetration depth and superfluid density are obtained, proving the presence of two superconducting \textit{s}-wave gaps.
\end{abstract}

\maketitle
%\section{\label{sec:level1}Introduction:}
\section{introduction}

Since the discovery of superconductivity in iron pnictides,\cite{ja800073m} their electronic phase diagrams, characterized by a close proximity of magnetic and superconducting phases, have been explored in great detail.
Of particular interest are the regions of the phase diagrams showing a crossover from magnetic order to superconductivity.
%In the vicinity of a magnetic quantum critical point, where long-range magnetic order is suppressed by varying a non-thermal control parameter such as external pressure or doping, superconductivity is often found.
Magnetic spin fluctuations, enhanced in the vicinity of a magnetic quantum critical point, can play an important role in the formation of Cooper pairs.
%It is commonly believed, that quantum critical points are often hidden by a superconducting state, so a suppression of the magnetic phase transition to zero temperature is not visible.
%In addition, magnetic spin fluctuations can play an important role in the formation of Cooper pairs.
In addition, the interplay of both magnetic order and superconductivity can lead to a phase with microscopic coexistence of both ground states.\cite{PhysRevB.89.054515, PhysRevB.80.100508, PhysRevB.82.014521, PhysRevB.81.174538, PhysRevB.81.140501}
As both states compete for the same electrons at the Fermi surface, the magnetic order parameter may be reduced below $T_c$.\cite{PhysRevB.82.014521, PhysRevB.81.174538, PhysRevB.81.140501}
In this work, we studied the system Ca$_{1-x}$Na$_x$Fe$_2$As$_2$.
The parent compound CaFe$_2$As$_2$ shows spin density wave order below the N\'{e}el-temperature $T_N~=~165$~K.\cite{PhysRevB.84.184534}
The magnetic phase transition is accompanied by a structural phase transition from a tetragonal to an orthorhombic structure.\cite{0953-8984-20-42-422201}
Increasing the Na amount in Ca$_{1-x}$Na$_x$Fe$_2$As$_2$, the magneto-structural phase transition is suppressed until it vanishes at a critical Na concentration \textit{x}$~\approx$~0.35.\cite{PhysRevB.84.184534}
Superconductivity is found for \textit{x}$~\geq$~0.3 with a maximum of $T_c~\approx$~34~K at an optimal doping of \textit{x}$~\approx$~0.66.
However, the interaction of superconductivity with the magnetic spin density wave in the region of 0.3$~\leq$~\textit{x}$~\leq$~0.35 has not been conclusively determined.

We studied the magneto-structural phase transition as well as both the superconducting and the magnetic order parameter and their interaction for different Na-substitution levels.
We find a suppression of the magneto-structural phase transition upon Na substitution.
For \textit{x}~=~0.35 and 0.50, microscopic coexistence of magnetic order and superconductivity is observed.
For the latter Na-substitution level, a reduction of the magnetic order parameter is observed below the superconducting transition temperature.
For \textit{x}~=~0.67, two superconducting gaps with \textit{s}-wave symmetry can be deduced from the temperature dependence of the superfluid density.

\section{experimental}
	\label{sec:experimental}
	
We examined mosaics of Ca$_{1-x}$Na$_x$Fe$_2$As$_2$ single crystals, which were grown by the self-flux technique as described by Johnsten \textit{et al}.\cite{PhysRevB.89.134507}
The samples were characterized by energy-dispersive X-ray spectroscopy (EDX), X-ray diffraction (XRD), susceptibility, magnetization, and specific-heat measurements.
The stoichiometry of the examined samples is \textit{x}~=~0.00, 0.35, 0.50, and 0.67 as determined by EDX.
A characterization of the magnetic properties was performed using SQUID magnetometry in large and small external magnetic fields.
%The XRD-measurements revealing \textit{c}--axis orientation of the platelets.
M\"ossbauer spectroscopy (MBS) experiments were performed in transmission geometry in a temperature range between 4.2 and 300~K using a CryoVac Konti IT cryostat.
As the $\gamma$ source, a $^{57}$Co in rhodium matrix was used.
 %with an emission line width of $\Gamma$~=~0.135(5)~mm/s was used.
The single crystals were aligned with the crystallographic \textit{c}--axis along the $\gamma$ direction.
To analyze the data, the hyperfine Hamiltonian including electric quadrupole and magnetic hyperfine interactions was diagonalized.
%using the \uppercase{M}\"o\uppercase{ssfit} software
In the paramagnetic temperature regime, the spectra were described by a doublet pattern, whereas in the magnetically ordered state a sextet pattern was used.
The magnetic order parameter is deduced from the $^{57}$Fe magnetic hyperfine field \textit{B}$_{\text{hf}}$.
The isomer shift $\delta$ is given with respect to $\alpha$-Fe.

Muon spin relaxation ($\mu$SR) experiments were performed at the $\pi$M3 and PiE1 beamlines of the Swiss Muon Source at the Paul Scherrer Institut, Switzerland, using the GPS and DOLLY spectrometers.
The single crystals were aligned with the crystallographic \textit{c}--axis along the muon beam.
Positively charged muons $\mu^+$, which are nearly 100\% spin-polarized due to parity violation during the pion decay, are implanted in the sample and thermalize at interstitial lattice sites, where they radioactively decay with a lifetime of 2.2~$\mu$s into two neutrinos and one positron.
As the muon decay involves the weak interaction, where parity conservation is violated, the positron is predominately emitted along the direction of the muon spin at the moment of the decay.
Measuring the time-resolved angular distribution of the emitted positrons allows to extract the time evolution of the muon spin polarization \textit{P}(\textit{t}).
%The \textit{z}-direction of the laboratory frame is determined by transverse field (TF) experiments.
The initial muon spin was rotated by approximately $-$45$^{\circ}$ with respect to the beam, which allows to measure the time evolution of the muon spin polarization $\perp$~\textit{c} and $\|$~\textit{c} by analysing the asymmetry of the up-down and forward-backward detector pair respectively.
If not stated otherwise, all presented measurements refer to the up-down detector pair.
The muon spin relaxation was measured for temperatures ranging from 1.6~K up to 300~K in zero field (ZF) and transverse magnetic fields (TF) up to 130~mT.
The $\mu$SR data were analyzed using the \uppercase{MUSRFIT} software package.\cite{Suter201269}

%As the $\mu^+$ has a spin 1/2, it can be used to study magnetic properties of the sample.
In a magnetically ordered material, the muon spin exhibits a Larmor precession with a frequency $\nu_{\mu}$, which is related to the local magnetic field \textit{B} at the muon site by $\nu_{\mu}$~=~\textit{B}$\gamma_{\mu}/(2\pi)$ (muon gyromagnetic ratio $\gamma_{\mu}=2\pi \times 135.5$~MHz/T).
The muon spin precession can be described in single crystals using the function\cite{0953-8984-9-43-002}
\begin{equation}
	\begin{split}
		P(t) = & \sum_{i=1}^N P(\nu_i) \left [ A_i \cos(\nu_i t) e^{-\lambda_T^i t} \right. \\ & \left. {}+ (1-A_i) e^{-\lambda_L^i t} \right],
		\end{split}
\end{equation}
with \textit{N} denoting the number of inequivalent muon sites contributing to the $\mu$SR signal, weighted by \textit{P}($\nu_i$).
In the case of 100\% magnetic ordering, $\sum{P(\nu_i)}=1$.
$A_i$ describes the oscillating part of the signal.
In contrast to powder samples, where \textit{A}$_i=2/3$ due to spatial averaging, for single crystals \textit{A}$_i \in [0,1]$.
$\lambda_T$, describing the damping of the oscillation, is a measure of the width of the static field distribution, also including dynamic contributions due to magnetic fluctuations.
The damping of the non-oscillating part, described by $\lambda_L$, is caused by dynamic magnetic fluctuations only.\cite{yaouanc2011muon}

To study the superconducting properties of the sample with $\mu$SR, a magnetic field $\mu_0 H_{\text{ext}}$ was applied parallel or perpendicular to the muon beam.
In type-II superconductors, a vortex lattice is formed for $\mu_0$\textit{H}$_{c1} <~\mu_0$\textit{H}$_{\text{ext}}<~\mu_0$\textit{H}$_{c2}$ resulting in a spatial magnetic field distribution.\cite{0953-8984-21-7-075701}
This magnetic field distribution causes an additional damping of the muon spin oscillation, which can be modeled via additional Gaussian terms of the form\cite{0953-8984-21-7-075701}
\begin{equation}
	P(t) = \sum_{i=1}^N{\left[ P_i  \cos(\gamma_{\mu} B_i t + \varphi)  e^{-\frac{1}{2} \sigma_i^2 t^2} \right ]  e^{-\frac{1}{2} \sigma_N^2 t^2}},
\end{equation}
where $\sigma_i$ describes the damping due to superconductivity, $\sigma_N$ the damping due to nuclear magnetic dipol contributions and $\varphi$ denotes the angle between the initial muon spin direction and the positron detector.
\\
The second moment $\braket{\Delta B^2}$ of the internal magnetic field distribution, \textit{n}(\textit{B}), is given by\cite{0953-8984-21-7-075701}
\begin{equation}
	\braket{\Delta B^2} = \sum_{i=1}^N \frac{A_i}{\sum_{i=1}^N A_i} \left [ \left( \frac{\sigma_i}{\gamma_{\mu}} \right )^2+ \left ( B_i - \braket{B} \right )^2 \right ]
	\label{eq:}
\end{equation} 
and related to the London penetration depths $\lambda$ by the relation\cite{PhysRevB.37.2349}
\begin{equation}
	\braket{\Delta B^2} = 0.00371 ~\Phi_0^2 \frac{1}{\lambda^4},
	\label{eq:brandt}
\end{equation} 
where $\Phi_0$ denotes the magnetic flux quantum.
$\lambda$ is related to the Cooper pair density by $n_s~\propto~1/\lambda^2$.

Therefore, $\mu$SR allows to independently measure the magnetic and superconducting order parameters via the determination of the zero field muon spin precession frequency $\nu_i$ and the London penetration depth, respectively.
Moreover, $\mu$SR is able to distinguish between non-magnetic and magnetic volume fractions in the sample.

%\begin{figure}[htbp]
	%\centering
		%\includegraphics[width=1.00\columnwidth]{MBS_all}
	%\caption{M\"ossbauer spectra of x=0.35, 0.50 and 0.67 for characteristic temperatures including the best fit.}
	%\label{fig:MBS_all}
%\end{figure}

\section{results and discussion}
	\label{sec:results}

\begin{figure}[h]
	\centering
		\includegraphics[width=0.90\columnwidth]{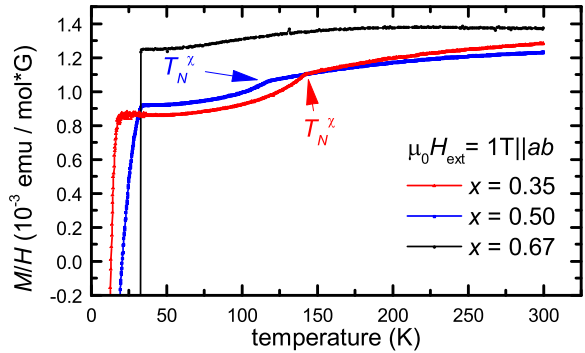}
	\caption{Temperature dependence of the static susceptibility $\chi~=~M/H$ of Ca$_{1-x}$Na$_{x}$Fe$_2$As$_2$ for \textit{x}~=~0.35, 0.50, and 0.67. The measurements were performed at an applied field of 1~T parallel to the \textit{ab}--plane. \textit{T}$_N^{\chi}$ denotes the magnetic transition temperature.}
	\label{fig:susceptibility1T}
\end{figure}

\begin{figure}[h]
	\centering
		\includegraphics[width=1.00\columnwidth]{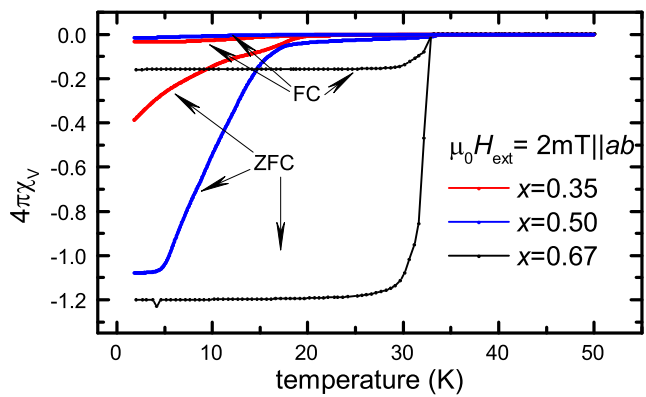}
	\caption{Temperature dependence of the magnetic susceptibility $\chi_v$ [zero (ZFC) and field cooled (FC)] of Ca$_{1-x}$Na$_{x}$Fe$_2$As$_2$ for \textit{x}~=~0.35, 0.50, and 0.67. The measurements were performed at an applied field of 2~mT parallel to the \textit{ab}--plane.}
	\label{fig:susceptibility20Oe}
\end{figure}

Measurements of the magnetic susceptibility of CaFe$_2$As$_2$ are reported by Harnagea \textit{et al}.\cite{PhysRevB.83.094523}
Magnetic susceptibility measurements in an applied field of 1~T parallel to the \textit{ab}--plane are shown in Fig.~\ref{fig:susceptibility1T}.
For \textit{x}~=~0.35, 0.50, and 0.67 a nearly linear decrease of the magnetic susceptibility is observed in the paramagnetic region, which is found also for many other iron pnictides.\cite{PhysRevB.85.224520,PhysRevB.81.024506, 0295-5075-86-3-37006}
The kink at 143~K, 119~K, and 150~K for \textit{x}~=~0.35, 0.50, and 0.67, respectively,  indicates the onset of the antiferromagnetic (AFM) ordering.
%For \textit{x}~=~0.67, a nearly linear decrease of the magnetic susceptibility is observed above the superconducting transition temperature $T_c$~=~34~K.
%The slope of this decrease changes at $\approx$~150~K.
%This change may be attributed to the onset of antiferromagnetic fluctuations.
Magnetic susceptibility measurements in an applied field of 2~mT parallel to the \textit{ab}--plane are shown in Fig.~\ref{fig:susceptibility20Oe}.
For \textit{x}~=~0.35, below 17~K a broad superconducting transition occurs.
As the diamagnetic shielding is not fully developed, only parts of the sample show superconductivity.
For \textit{x}~=~0.50, a two-step superconducting transition occurs.
%Below 34~K, small parts of the sample volume show a superconducting response resulting in a slightly negative susceptibility.
Below 34~K, a slightly negative magnetic susceptibility is measured, which indicates a superconducting phase in a small volume of the sample.
Below 17~K, a broad second transition occurs, where bulk superconductivity is formed resulting in the full superconducting response.
For the further treatment of the \textit{x}~=~0.50 sample, 17~K will be considered as the superconducting transition temperature.
This two-step behavior as well as the broad transition indicates an inhomogeneous sample.
The sample with \textit{x}~=~0.67 shows bulk superconductivity below 34~K.
The sharp transition indicates a homogeneous sample.

\subsection{The parent compound CaFe$_2$As$_2$}
\begin{figure}[htbp]
	\centering
		\includegraphics[width=1.00\columnwidth]{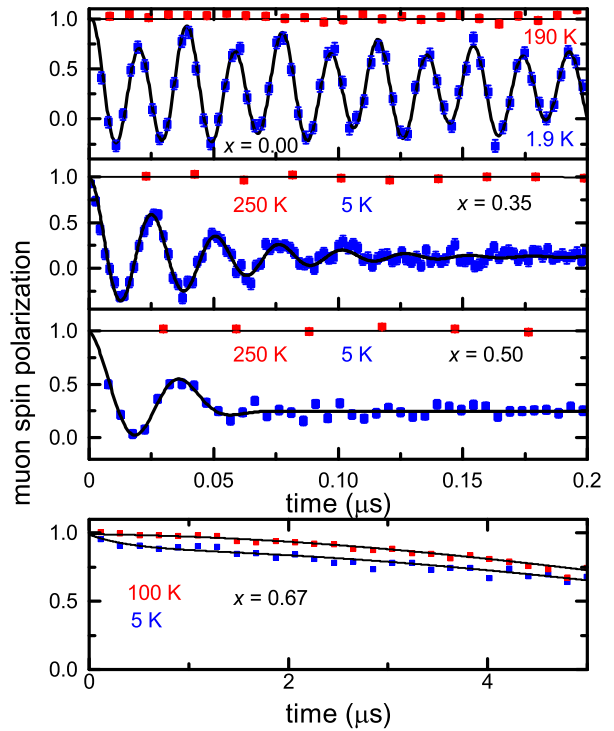}
	\caption{ZF-$\mu$SR time spectra of Ca$_{1-x}$Na$_{x}$Fe$_2$As$_2$ for characteristic temperatures in the paramagnetic and magnetically ordered state. The well-defined muon spin precession for \textit{x}~=~0.00, 0.35, and 0.50 indicate long-range commensurate magnetic order. For \textit{x}~=~0.67, non-magnetic behavior is found down to lowest measured temperatures.}
	\label{fig:muSR_all_neu}
\end{figure}

\begin{figure}[htbp]
	\centering
		\includegraphics[width=1.00\columnwidth]{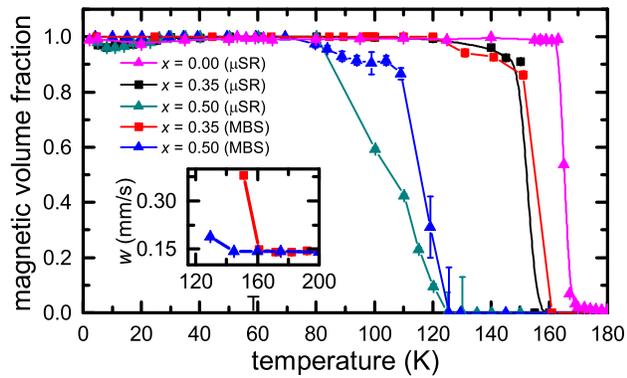}
	\caption{Magnetic volume fraction as a function of temperature (lines are guides to the eye only) of Ca$_{1-x}$Na$_{x}$Fe$_2$As$_2$ obtained from ZF-$\mu$SR and M\"ossbauer spectroscopy for \textit{x}~= 0.00, 0.35, and 0.50. The inset shows the temperature dependence of the M\"ossbauer linewidth \textit{w}. The abrupt increase of \textit{w} indicates magnetic ordering.}
	\label{fig:MVF_all}
\end{figure}

\begin{figure*}[htb]
	\centering
	\captionsetup[subfloat]{position=top, parskip=0pt, aboveskip=0pt, justification=raggedright, labelseparator=none, farskip=0pt, nearskip=2pt, margin=0pt, captionskip=-15pt, font=normalsize, singlelinecheck=false}
	\subfloat[]{\includegraphics[width= 8.6cm]{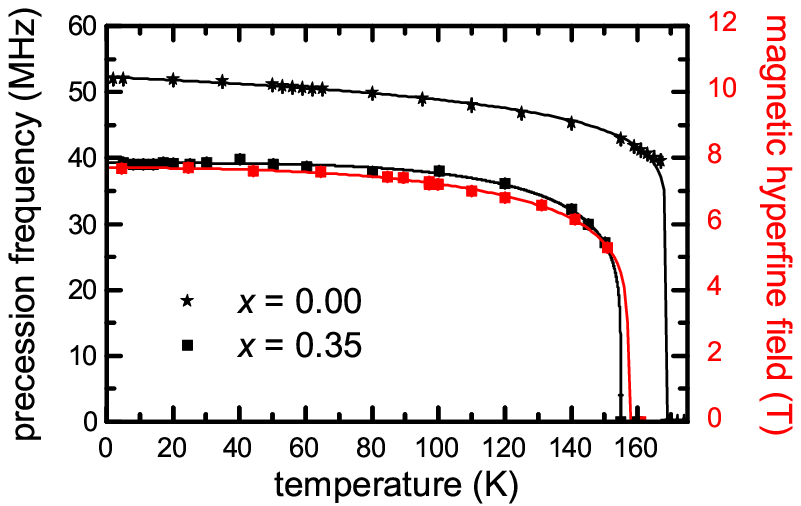} \label{fig:SE2858muSR_OP+Freq}}\quad
	\subfloat[]{\includegraphics[width= 8.6cm]{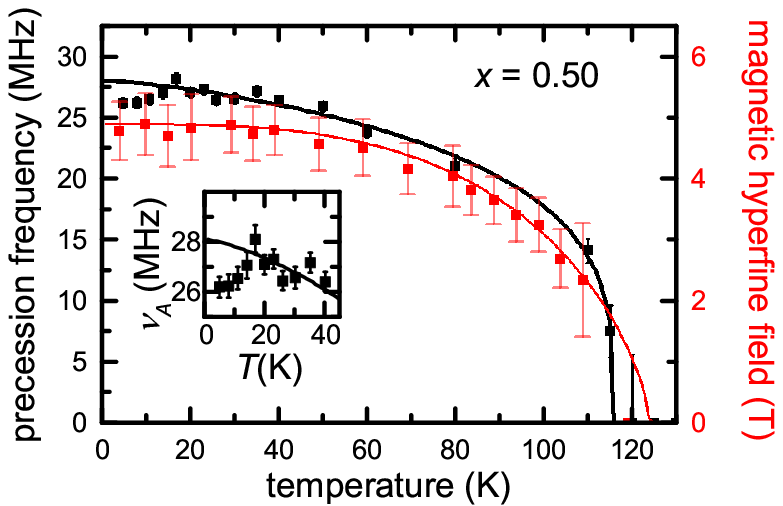} \label{fig:SE2997muSR_OP+Freq}} 
	\caption{Temperature dependence of the magnetic order parameter for (a) \textit{x}~=~0.00 and 0.35 and (b) \textit{x}~=~0.50 including best order parameter fits according to Eq.~(\ref{eq:OP}). The  magnetic hyperfine field for \textit{x}~=~0.50 is Gaussian-distributed. The average value of the Gaussian distribution is shown with one standard deviation as error bar. The inset in (b) shows the low temperature regime, where superconductivity occurs below \textit{T}$_c$~=~17~K.}
	\label{fig:OPT}
\end{figure*}

The time evolution of the muon spin polarization in ZF is shown in Fig.~\ref{fig:muSR_all_neu}.
In the paramagnetic temperature regime, a weak Gauss-Kubo-Toyabe damping of the signal is observed caused by the dipole-dipole interaction of the muon magnetic moment with randomly oriented nuclear magnetic moments.
The temperature dependence of the magnetic volume fraction (MVF) is shown in Fig.~\ref{fig:MVF_all}.
To describe MVF as a function of \textit{T}, two temperatures are defined: \textit{T}$_N^{\text{onset}}$ describes the highest temperature with a finite MVF and \textit{T}$_N^{100~\%}$ describes the highest temperatures with MVF~=~100~\%~$\hat{=}$~1.
The sharp transition between \textit{T}$_N^{\text{onset}}$~=~167~K and \textit{T}$_N^{100~\%}$~=~163~K indicates a homogeneous sample.
Below 167~K, two magnetically inequivalent muon stopping sites \textit{A} and \textit{B} with a temperature-independent occupation ratio of \textit{P}$_A$:\textit{P}$_B$=~80:20 were observed as it was found in BaFe$_2$As$_2$.\cite{PhysRevB.78.214503}
The signal fraction corresponding to muons stopping at site \textit{A} show a well-defined sinusoidal oscillation below 167~K.
This indicates static long-range commensurate magnetic order.
The temperature dependence of $\nu_A$ is shown in Fig.~\ref{fig:OPT}.
The step-like behavior indicate a first-order transition, as it was seen in SrFe$_2$As$_2$. \cite{PhysRevB.78.180504}
The signal fraction corresponding to muons stopping at site \textit{B} show an exponential relaxation below 167~K and a well-defined sinusoidal oscillation below 60~K.
This indicates a a broad field distribution at temperatures between 60~K and 167~K at the muon stopping site \textit{B} suppressing a coherent oscillation of the muon spins, which is contrary to the observations in BaFe$_2$As$_2$, where the two oscillation frequencies were obtained at all temperatures below $T_N$,\cite{PhysRevB.78.214503} but consistent with LaOFeAs.\cite{PhysRevLett.101.077005}
Below 60~K, the two precession frequencies have a temperature-independent ratio of $\nu_A$/$\nu_B~\approx$~1.9.
Therefore, the magnetic field at the muon stopping site \textit{A} is higher than on site \textit{B}.
Site \textit{A} is located next to the FeAs-layer.\cite{Amato2009606}
The smaller value of the magnetic field at site \textit{B} indicates a muon stopping site more distant from the FeAs layer.
As the precession frequency is proportional to the magnetic field at the muon site, this ratio is different from BaFe$_2$As$_2$\cite{PhysRevLett.107.237001} and SrFe$_2$As$_2$,\cite{PhysRevB.78.180504} showing ratios of 4.1 and 3.4.
The ionic radii of the alkaline earth metals scale like Ca~$<$~Sr~$<$~Ba.\cite{a12967}
As a consequence, the crystallographic \textit{c}--axis is shortest for CaFe$_2$As$_2$\cite{PhysRevB.84.184534} and longest for BaFe$_2$As$_2$.\cite{ANIE200803641}
%The muon stopping site \textit{B}, showing the smaller precession frequency $\nu_B$, is located next to the alkaline earth metals.
By shrinking the crystallographic \textit{c}--axis, the distance between site \textit{B} and the ordered iron magnetic moments in the FeAs-layer is reduced.
The muon spin interacts with the ordered electronic moments via dipole-dipole and transferred Fermi-contact interaction.
Both interactions are sensitive to the distance between the muon spin and the iron ordered moments.
This implies that the change of the frequency ratio in the undoped compounds has a structural origin.

M\"ossbauer measurements by Alzamora \textit{et al}.\cite{0953-8984-23-14-145701} on CaFe$_2$As$_2$ show a first-order-like magneto-structural phase transition below 173~K.
They report a saturated magnetic hyperfine field of $\approx$~10.1~T at 4.2~K.
The angle between the magnetic hyperfine field and the principal axis of the electric field gradient was reported as 94(4)$^{\circ}$.
Therefore, as the principal axis of the electric field gradient is parallel to the crystallographic \textit{c}--axis,\cite{0953-8984-23-14-145701} the magnetic hyperfine field is located in the \textit{ab}--plane.\cite{0953-8984-23-14-145701, PhysRevB.78.100506}
These M\"ossbauer results are consistent with our $\mu$SR results showing that the magnetic moments are located in the \textit{ab}--plane.

\subsection{Underdoped Ca$_{0.65}$Na$_{0.35}$Fe$_2$As$_2$ and Ca$_{0.50}$Na$_{0.50}$Fe$_2$As$_2$}
\begin{figure}[htbp]
	\centering
		\includegraphics[width=1.00\columnwidth]{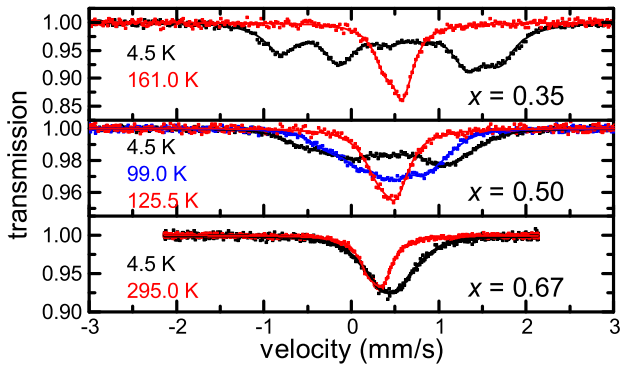}
	\caption{M\"ossbauer spectra of Ca$_{1-x}$Na$_{x}$Fe$_2$As$_2$ for characteristic temperatures in the paramagnetic and magnetically ordered state. The sextet structure of the M\"ossbauer spectra for \textit{x}~=~0.35 and 0.50 indicate long-range commensurate magnetic order. For \textit{x}~=~0.67, a broadening below $\approx$~150~K is observed, which cannot be described by a change in $V_{zz}$ indicating small magnetic fields at the iron nucleus.}
	\label{fig:MBS_all}
\end{figure}

\begin{figure}[htbp]
	\centering
		\includegraphics[width=1.00\columnwidth]{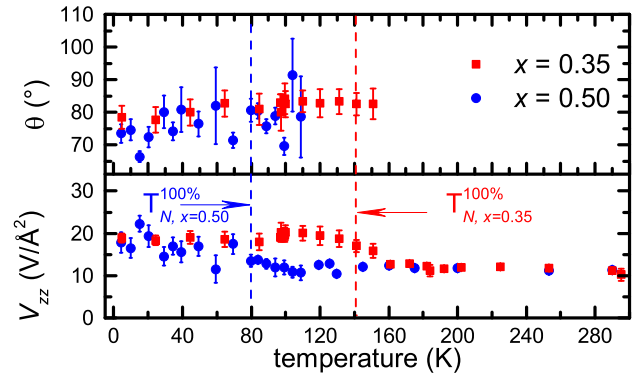}
	\caption{Temperature dependence of $\theta$, the polar angle between the magnetic hyperfine field \textit{B}$_{\text{hf}}$ and the principal axis of the electric field gradient, \textit{V}$_{zz}$. The increase of \textit{V}$_{zz}$ indicates a change in the electromagnetic environment of the $^{57}$Fe-nucleus and corresponds to the magneto-structural phase transition. The vertical dashed lines denotes the highest temperature, where the samples shows a MVF of 100~\%.}
	\label{fig:VzzTheta}
\end{figure}

M\"ossbauer spectra for characteristic temperatures in the paramagnetic and magnetically ordered states are shown in Fig.~\ref{fig:MBS_all}.
In the paramagnetic state, an asymmetric doublet structure, which is caused by the interaction of the nucleus with an electric field gradient (EFG), was observed for both stoichiometries.
However, in the principal-axis system, the EFG is fully determined by its \textit{z}-component \textit{V}$_{zz}$ and the asymmetry parameter $\eta$ (which turned out to be zero for all investigated temperatures).
At room temperature, for \textit{x}~=~0.35 and 0.50 a value of \textit{V}$_{zz}=~11.2(5)$~V/\AA$^2$ was obtained.
With decreasing temperature, \textit{V}$_{zz}$ slightly increases to 13.0(5)~V/\AA$^2$ directly above the magnetic transition temperature as shown in Fig.~\ref{fig:VzzTheta}.
With the onset of magnetic order, \textit{V}$_{zz}$ increases to 19(2)~V/\AA$^2$ and 18(3)~V/\AA$^2$ for \textit{x}~=~0.35 and 0.50, respectively, and remains constants within error bars down to lowest temperature.
This increase can be assigned to a magneto-structural phase transition as observed for other 122 compounds.\cite{1367-2630-11-2-025023, PhysRevLett.101.257003}
%Furthermore, in the magnetically ordered state, angles of $\varphi^{x=0.35}_{\text{AFM}}=~17.7(1.2)^{\circ}$ and $\varphi^{x=0.50}_{\text{AFM}}=~18.3(8)^{\circ}$ were obtained.
%This indicates a turning of the principal axis of the EFG towards the crystallographic \textit{c}--axis.
Furthermore, the principal axis of the EFG is parallel to the crystallographic \textit{c}--axis.
%, as it was also determined for the parent compound.\cite{0953-8984-23-14-145701}
The temperature dependence of the angle $\theta$ between the principal axis of the EFG and the magnetic hyperfine field, obtained from a fit to the data, is shown in Fig.~\ref{fig:VzzTheta}.
We have obtained $\theta$~=~$80(5)^{\circ}$ and $71(5)^{\circ}$ for \textit{x}~=~0.35 and 0.50, respectively, in the fully ordered state were obtained.
These values indicate a tilting of the magnetic moments out of the \textit{ab}--plane.

The time evolution of the muon spin polarization in ZF on Ca$_{1-x}$Na$_{x}$Fe$_2$As$_2$ (with \textit{x}~=~0.00, 0.35, 0.50, and 0.67) is shown in Fig.~\ref{fig:muSR_all_neu}.
As for the parent compound described above, a weak Gauss-Kubo-Toyabe damping of the signal is observed caused by the dipole-dipole interaction of the muon magnetic moment with randomly oriented nuclear magnetic moments in the paramagnetic temperature regime.
The onset of long-range commensurate magnetic ordering below \textit{T}$_N^{\text{onset}}$~=~160(2)~K and 125(3)~K for \textit{x}~=~0.35 and 0.50, respectively, is indicated by the appearance of a well-defined muon spin precession with two frequencies $\nu_A$ and $\nu_B$.
The occupation probability is independent of temperature and the Na-substitution level (\textit{P}($\nu_A$):\textit{P}($\nu_B$)=~80:20).
We conclude that two magnetically inequivalent muon sites are present in a homogeneous magnetically ordered phase.
The temperature dependence of the muon frequency $\nu_A$ for both samples is shown in Fig. \ref{fig:OPT}(a) and \ref{fig:OPT}(b).
The onset temperature is consistent with the results of the macroscopic magnetizations measurements by SQUID magnetometry.
The two frequencies have a temperature-independent ratio of $\nu_A$/$\nu_B \approx$~4 and 8 for \textit{x}~=~0.35 and 0.50, respectively, which are different compared to the undoped compound showing $\nu_A$/$\nu_B \approx$~1.9.
%This change of the frequency ratio upon doping, compared to the undoped compound showing $\nu_A$/$\nu_B~\approx$~1.9 indicates a change of the magnetic structure.
This change is more drastic compared to other 122 compounds.
BaFe$_2$As$_2$ shows a ratio of 4.1,\cite{PhysRevLett.107.237001} which increases to 4.47 for Ba$_{0.77}$K$_{0.23}$Fe$_2$As$_2$\cite{PhysRevLett.107.237001} and 4.5 for Ba$_{0.7}$Na$_{0.3}$Fe$_2$As$_2$.\cite{Maeter}
SrFe$_2$As$_2$ shows a ratio of 3.4,\cite{PhysRevB.78.180504} which changes to 3.6 for Sr$_{0.5}$Na$_{0.5}$Fe$_2$As$_2$.\cite{PhysRevB.80.024508}
Substituting an alkaline-earth metal by an alkaline metal with a smaller ionic radius (Ba$\rightarrow$Na, Sr$\rightarrow$Na), leads to a smaller change in the \textit{c}--axis parameter and the change of the precession frequency ratio $\nu_A$/$\nu_B$ occurs at lower substitution levels than by substituting with an alkaline metal with a larger ionic radius (Ca$\rightarrow$Na, Ba$\rightarrow$K, Sr$\rightarrow$K).\cite{Maeter, PhysRevB.78.180504, PhysRevB.80.024508, PhysRevB.84.184534,ANIE200803641, PhysRevB.88.094510,cm1028244}
The ionic radii scale like Ca~$<$~Na~$<$~Sr~$<$~Ba~$<$~K.\cite{a12967}
By increasing the Na-substitution level in Ca$_{1-x}$Na$_x$Fe$_2$As$_2$, the crystallographic \textit{c}--axis is elongated resulting in an increased distance of the muon spin at the site \textit{B} from the iron ordered moments in the FeAs-plane.
In addition, the small tilting of the magnetic moments out of the \textit{ab}--plane may also change the magnetic field at the muon site.
However, a change in the Fermi-contact interaction for both muon stopping sites cannot be ruled out.
Additionally, a signal fraction in the forward-backward detector pair (not shown here) showing an exponential relaxation is observed in the magnetically ordered phase.
This signal fraction increases as a function of Na-substitution level.
This is consistent with the tilting of the magnetic moments out of the \textit{ab}--plane as observed by M\"ossbauer spectroscopy.

The temperature dependence of the magnetic volume fraction was determined independently by ZF- and TF-$\mu$SR measurements as well as M\"ossbauer spectroscopy.
TF-$\mu$SR experiments were performed by applying a magnetic field of 5~mT perpendicular to the initial muon spin polarization.
The temperature dependence of the MVF is shown in Fig.~\ref{fig:MVF_all}.
%Magnetic ordering  is found below \textit{T}~=~161~K and \textit{T}~=~125~K for \textit{x}~=~0.35 and 0.50.
The onset of the magnetic ordering is also indicated by the abrupt increase of the M\"ossbauer line width \textit{w} due to appearance of a magnetic hyperfine field.
This increase is shown in the inset of Fig.~\ref{fig:MVF_all} leading to \textit{T}$_{\text{onset}}~=$~161(2)~K and 125(3)~K for \textit{x}~=~0.35 and 0.50, respectively.
Therefore, the obtained characteristic temperatures of the magnetic phase transition are of equal value within error bars for both $\mu$SR and MBS and coincide with the magnetic phase transition temperature obtained by magnetic susceptibility measurements, \textit{T}$_N^{\chi}$.
The temperature width of the phase transition, $\Delta$\textit{T} = \textit{T}$_N^{\text{onset}}$ $-$\textit{T}$_N^{100~\%}$, increased from 4~K to 21~K and 45~K for \textit{x}~=~0.00, 0.35, and 0.50, respectively.
We attribute this increased width of the magnetic phase transition to the increased degree of disorder due to the Na-substitution.
%Within a pure paramagnetic model, the onset of magnetic ordering is indicated by the abrupt increase of the line width \textit{w}, which occurs below \textit{T}$_{\text{onset}}^{\text{x=0.35}}=$~161~K and \textit{T}$_{\text{onset}}^{\text{x=0.50}}=$~125~K, respectively, as shown in the inset of Fig.~\ref{fig:MVF_all}.
The MVF is constant below \textit{T}$_N^{100~\%}$~= 163(2)~K, 140(3)K and 80(3)K for \textit{x}~=~0.00, 0.35, and 0.50, respectively, proving bulk magnetic order.
%The sharp transition in both, $\mu$SR and M\"ossbauer spectroscopy, for \textit{x}~=~0.35 indicates a homogeneous sample, while the broad transition for \textit{x}~=~0.50 indicates an inhomogeneous sample.
%A nearly constant MVF of $\approx 100$~\% below \textit{T}$^{x=0.35}_{100\%}=~150$~K and \textit{T}$^{x=0.50}_{100\%}=~80$~K shows bulk magnetism.
In the magnetically ordered state, a well-resolved sextet was observed in the M\"ossbauer spectra for \textit{x}~=~0.35, as it is shown in Fig.~\ref{fig:MBS_all} at \textit{T}~=~4.5~K.
This proves a static commensurate magnetic ground state with a well-defined hyperfine field.
For \textit{x}~=~0.50, the sextet is less clearly resolved and the spectra were modeled using a Gaussian distribution of magnetic hyperfine fields.
This takes into account a higher degree of disorder than in the sample with \textit{x}~=~0.35.
Consistently also the $\mu$SR transverse relaxation rate $\lambda^{\text{T}}$ is largest for \textit{x}~=~0.50 as can be seen by a much faster suppression of the ZF oscillation in Fig.~\ref{fig:muSR_all_neu} compared to lower Na-substitution levels.
%This is also indicated by the rapid suppression of the muon precession as shown in Fig.~\ref{fig:muSR_all_neu}, corresponding to a large transversal relaxation rate, $\lambda_{\text{T}}$, which is a measure for the static width of the field distribution.
The temperature dependence of the obtained magnetic hyperfine field of \textit{x}~=~0.35 is shown in Fig.~\ref{fig:SE2858muSR_OP+Freq}.
%In Fig.~\ref{fig:SE2997muSR_OP+Freq} the mean value of the Gaussian distributed magnetic hyperfine field of \textit{x}~=~0.50 is shown with one standard deviation as errorbars denote [2 ln(2)]$^{-0.5} \cdot$HWHM$_{\text{Gauss}}$.
The temperature dependence of the mean value of the obtained Gaussian distributed magnetic hyperfine field \textit{B}$_{\text{hf}}$ for \textit{x}~=~0.50 is shown in Fig.~\ref{fig:SE2997muSR_OP+Freq}.
%The error bars are conncted to the standard deviation over the relation [2 ln(2)]$^{-0.5} \cdot$HWHM$_{\text{Gauss}}$.

Both $\nu_A$ and \textit{B}$_{\text{hf}}$ were analyzed using a fit to the temperature-dependent order parameter (\textit{M}) of the form
\begin{equation}
	M(T) = M(T~=~0)  \left [1 - \left ( \frac{T}{T_N} \right )^\alpha \right ]^\beta
	\label{eq:OP}
\end{equation}
for \textit{x}~=~0.35 and 0.50, respectively, above the superconducting transition temperatures (17~K in both cases).
For x~=~0.35, a fit to Eq.(\ref{eq:OP}) above \textit{T}$_c$ for both $\nu_A$(\textit{T}) and \textit{B}$_{\text{hf}}$(\textit{T}) represents the data in the whole temperature range, i.e., also below the superconducting transition.
Therefore, no interaction between the magnetic and superconducting order parameter is detectable.
For \textit{x}~=~0.50, the muon frequency is reduced by approximately 7~\% below 17~K while the magnetic hyperfine field shows no reduction and is well-represented by Eq.(\ref{eq:OP}).
This reduction of the muon precession frequency proves the microscopic coexistence of magnetic order and superconductivity and their competition.
Possibly \textit{B}$_{\text{hf}}$(\textit{T}) shows no signatures for a reduction below 17~K, since the reduction of 7~\% is within the hyperfine field error bars.
Alternatively, the muon precession frequency may be reduced due to a spin reorientation below $T_c$ rather than a reduction of the magnetic order parameter.
However, in this scenario the angle $\theta$ between the magnetic hyperfine field \textit{B}$_{\text{hf}}$ and the principal axis of the EFG would change.
The temperature dependence of $\theta$ is shown in Fig.~\ref{fig:VzzTheta}.
Since there is no systematic change of $\theta$ below $T_c$ observed within error bars, a significant spin reorientation can be ruled out.
Both order parameters are weakly coupled compared to Ba$_{1-x}$Na$_{x}$Fe$_2$As$_2$,\cite{Maeter} where a reduction of the muon spin frequency and therefore of the magnetic order parameter of $\approx$~65~\% appears.
\begin{figure}[htbp]
	\centering
		\includegraphics[width=1.00\columnwidth]{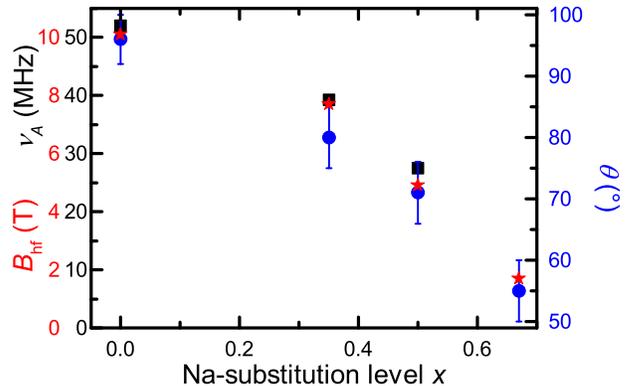}
	\caption{Low-temperature saturation values of the magnetic hyperfine field \textit{B}$_{\text{hf}}$, the muon spin precession frequency $\nu_A$ and $\theta$, the angle between \textit{B}$_{\text{hf}}$ and the principal-axis of the EFG, as a function of the Na-substitution level. Upon Na substitution, an reduction of the magnetic hyperfine field and the muon spin precession frequency is observed proving a reduction of the magnetic order parameter. The decrease $\theta$ indicate a tilting of the magnetic moments out of the \textit{ab}--plane as a function of the Na-substitution level.}
	\label{fig:PN_OP}
\end{figure}

%Again, the transition is broader in the \textit{x}~=~0.50 sample, which is consistent with other data.
%Small deviations from 100~\% can be attributed to muons that stop outside the sample, for example in the sample holder or cryostat walls.
%The begin of the magneticaly ordering observed by $\mu$SR and M\"ossbauer specrtroscopy is in good agreement with the transition temperatures obtained by susceptibility measurements with 1~T$\parallel ab$-plane, which are shown in Fig.~\ref{fig:susceptibility1T}.
%Also, susceptibility measurements with 20~Oe$\parallel ab$-plane indicating bulk superconductivity for 

%\begin{figure}[htbp]
	%\centering
		%\includegraphics[width=1.00\columnwidth]{pinning}
	%\caption{$\mu$SR pinning experiment for Ca$_{0.65}$Na$_{0.35}$Fe$_2$As$_2$ (upper panel) and Ca$_{0.5}$Na$_{0.5}$Fe$_2$As$_2$ (lower panel)at 5~K. The field was isothermally changed from 500~mT to 450~mT. As nearly 100~\% of the signal follows the external field, bulk superconductivity can be exluded from $\mu$SR point of view.}
	%\label{fig:pinning}
%\end{figure}

\subsection{Landau theory for coupled order parameters}
\begin{figure*}[htb]
	\centering
		\includegraphics[width=0.90\textwidth]{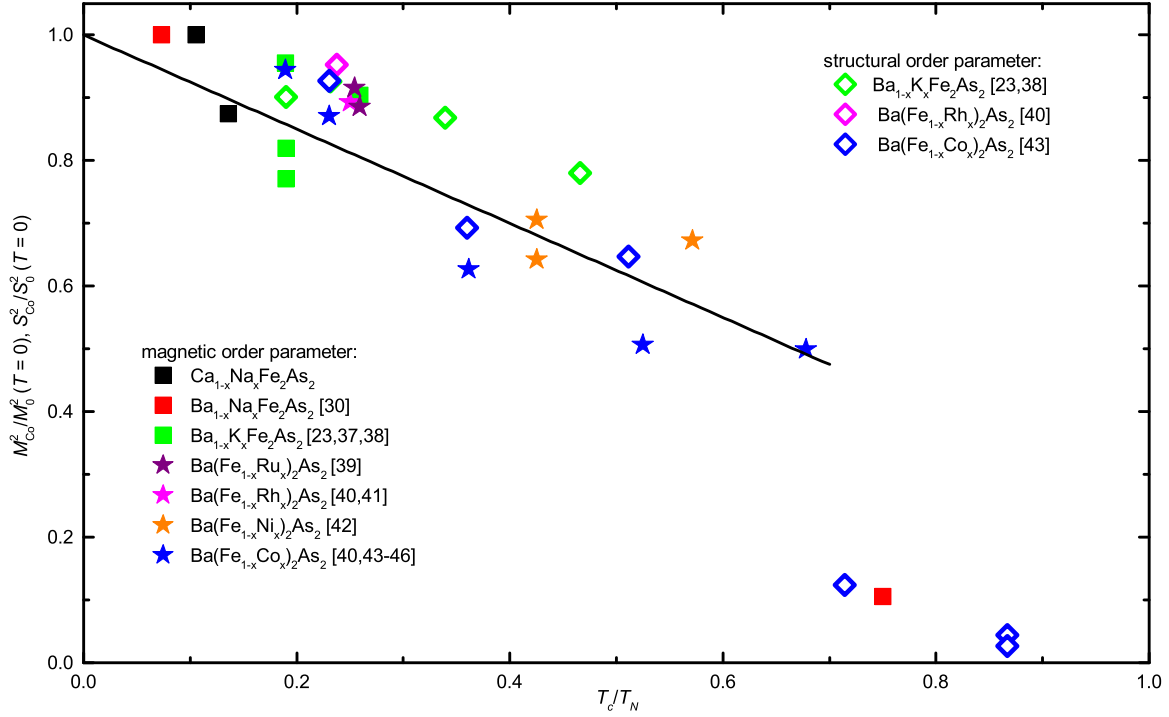}
	\caption{$|\vec{M}_{\text{co}}|^2/|\vec{M}_0|^2(T=0)$ as a function of $T_c/T_N$ for various 122 compounds.\cite{Maeter, PhysRevLett.107.237001, PhysRevB.85.144506, PhysRevB.83.172503, PhysRevB.83.054514, PhysRevB.81.134512, PhysRevB.84.024509, PhysRevLett.108.247002, PhysRevB.81.134512, PhysRevLett.104.057006, PhysRevLett.105.057001, PhysRevLett.103.087002, PhysRevLett.106.257001, PhysRevLett.107.237001} $|\vec{M}_{\text{co}}|^2$ denotes the magnetic order parameter in the region of coexistence with superconductivity. $|\vec{M}_0|^2$ denotes the magnetic order parameter without any superconductivity. The reduction of the magnetic order parameter in the coexistence region increases with increase of $T_c/T_N$ following Eq.~(\ref{eq:OPReductionT0}), the best linear fit fot $T~<~0.7$ is shown as the solid line. }
	\label{fig:OPReduction}
\end{figure*}

In the case of coexistence of magnetic order and superconductivity, an interaction between both order parameters is expected to be present.
This may change the magnitude of the order parameters and alter the critical temperatures with respect to the decoupled situation.
%Coexistence of magnetic order and superconductivity is discovered in various 122 compounds \textbf{refs}, but the magnitude of the reduction of the magnetic order parameter, if existend, is poorly studied.
%For coexistence 
%In the region of coexistence of magnetic order and superconductivity in the phase diagram, changes of the orde para
We used a Landau theory to describe the coupling between the superconducting order parameter $\psi(\vec{r})$ and the magnetic order parameter  $\vec{M}(\vec{r})$ in the case of coexistence and to describe the dependence of the reduction of the magnetic order parameter on the critical temperatures.
The homogeneous free-energy functional in the absence of an external field is given by \cite{Suhl1978225, 2221370109, PhysRevB.82.014521, 0953-2048-23-5-054011}
\begin{align}
	F[\psi, \vec{M}] = \int d^3r& \Bigg [ \frac{\alpha}{2} |\psi(\vec{r})|^2 + \frac{\beta}{4} |\psi(\vec{r})|^4  + \frac{a}{2} |\vec{M}(\vec{r})|^2\notag \\
	&+ \frac{b}{4} |\vec{M}(\vec{r})|^4 + \frac{d}{2}|\psi(\vec{r})|^2|\vec{M}(\vec{r})|^2    \Bigg ],
	\label{eq:GL}
\end{align}
where the coupling between of the two order parameters is contained in the last term.
Following the common approach, $\alpha$ and \textit{a} are described as
\begin{align}
	a =& a_0 (T - T_{N0}), \label{eq:GLa}\\
	\alpha =& \alpha_0 (T - T_{c0}), \label{eq:GLalpha}
\end{align}
where $T_{N0}$ and $T_{c0}$ denote the bare magnetic and superconducting transition temperatures, respectively.
The bare transition temperatures describe the decoupled case.
%The measured transition temperature may be reduced due to the order parameter competition.
To ensure that the free-energy has a minimum and an that the two order parameters compete with each other, $\beta~>~0$, $b~>~0$, and $d~>~0$ are required.
The two order parameters are obtained by minimizing the free-energy functional $F[\psi, \vec{M}]$.
The order parameters in the coexistence region are then given by
\begin{align}
	|\psi|^2 =& -\frac{\alpha b - a d}{b \beta - d^2}, \quad  &\alpha b - a d ~<~ 0, \label{eq:GLSolvedPsi} \\
	|\vec{M}|^2 =& -\frac{a \beta - \alpha d}{b \beta - d^2},\quad  &a \beta - \alpha d~<~ 0.
\end{align}
This solution is stable for a sufficiently small coupling \textit{d}, which satisfies 
$b \beta - d^2~>~0$.
For the further treatment, $T_{c0}~<~T_{N0}$ is assumed, which is the case for sufficiently low doping in the 122 compounds.
This results in $T_{N0}~=~T_N$.
At $T_{c}$, where $|\psi|^2=~0$, Eq.~(\ref{eq:GLSolvedPsi}) reduces to 
\begin{equation}
	|\psi(T_{c})|^2= 0 = \alpha_0 b (T_c - T_{c0}) - a_0 d (T_{c} - T_{N})
\end{equation}
which results in a linear relation between the measured and bare superconducting transition temperatures.
The temperature dependence of the magnetic order parameter in the coexistence regime is then given by 
\begin{align}
	\vec{M}_{\text{co}}^2(T) &= \frac{1}{b \beta - d^2} \bigg \{[\alpha_0 d-a_0 \beta] T\notag \\ & + \left [a_0 \beta -  \frac{a_0}{b}d^2  \right ]T_{\text{N}}- \left [ 1 - \frac{a_0 d}{\alpha_0 b} \right] T_{\text{c}} \bigg\}. \label{eq:GLMCo}
\end{align}
%Without any coupling (\textit{d}~=~0), the result of the pure magnetic order parameter $|\vec{M}_0|^2(T)~=~-a/b~=~a_0/b\cdot[T_N-T]$ is obtained.
To investigate the reduction of the magnetic order parameter below the superconducting transition temperature, the ratio of $|\vec{M}_{\text{co}}|^2(T)$ and $|\vec{M}_0|^2(T)~=~a_0 (T_N - T)/b$ for $T~<~T_c$ is calculated.
The reduction is maximal for $T~\rightarrow~0$.
The ratio is then given by
\begin{equation}
	\frac{\vec{M}_{\text{co}}^2}{\vec{M}_0^2} (T=0) = 1 - \frac{d}{a_0} \frac{\alpha_0 b - a_0 d}{b \beta - d^2} \frac{T_{\text{c}}}{T_{\text{N}}}.
\label{eq:OPReductionT0}
\end{equation}
%Therefore, for a given coupling \textit{d}, the magnetic order parameter is reduced proportional to $T_c/T_N$.
Literature data for $|\vec{M}_{\text{co}}|^2/|\vec{M}_0|^2(T=0)$ as a function of $T_c/T_N$ for various 122 compounds, measured by neutron scattering and $\mu$SR, are shown in Fig.~\ref{fig:OPReduction}.\cite{Maeter, PhysRevLett.107.237001, PhysRevB.85.144506, PhysRevB.83.172503, PhysRevB.83.054514, PhysRevB.81.134512, PhysRevB.84.024509, PhysRevLett.108.247002, PhysRevB.81.134512, PhysRevLett.104.057006, PhysRevLett.105.057001, PhysRevLett.103.087002, PhysRevLett.106.257001, PhysRevLett.107.237001}
In addition, the reduction of the structural order parameter $S~=~(a-b)/(a+b)$ is plotted in Fig.~\ref{fig:OPReduction}.
The magnetic and structural order parameters show the same behavior, as they are strongly coupled in the 122 pnictides.\cite{PhysRevB.89.144511,Yildirim2009425, PhysRevLett.101.257003, PhysRevB.78.180504, PhysRevB.78.100506, PhysRevB.79.220511, PhysRevB.83.054503, PhysRevB.80.094504}
$|\vec{M}_{\text{co}}|^2/|\vec{M}_0|^2(T=0)$ shows a nearly linear decrease as a function of $T_c/T_N$ and hence the reduction of the magnetic order parameter shows a square root behavior.
This corresponds to a constant slope in Eq.~(\ref{eq:OPReductionT0}), which implies a similar coupling strength of the magnetic and superconducting order parameters in the 122 compounds.
For $T_c/T_N \ll~1$, experimental data show no reduction of the magnetic order parameter in the coexistence regime.
This may result from the fact that the samples show SC only in parts of the sample volume, which has been observed for, e.g., Ca$_{0.65}$Na$_{0.35}$Fe$_2$As$_2$.
Equation~(\ref{eq:OPReductionT0}) qualitatively describes the reduction of $|\vec{M}_{\text{co}}|^2/|\vec{M}_0|^2(T=0)$ with increasing $T_c/T_N$ ratio for 0~$<$~$T_c/T_N$~$<$~0.7.
The systematic deviations for $T_c/T_N$~$>$~0.7 indicate an increase in the coupling strength.
This means that the coupling is more effective if $T_c~\approx~T_N$.
%, which implies, that the coupling of the order parameter in the 122 compounds is of equal value and the ratio $T_c/T_N$ determines the magnitude of the reduction of the magnetic order parameter in the coexistence region.

%\color[rgb]{1,0,0}Fuers Bild zum einfuegen am Ende: Data taken from Refs. Ba$_{1-x}$Na$_x$Fe$_2$As$_2$ \cite{Maeter}, Ba$_{1-x}$K$_x$Fe$_2$As$_2$ \cite{PhysRevLett.107.237001, PhysRevB.85.144506, PhysRevB.83.172503}, Ba(Fe$_{1-x}$Ru$_{x}$)$_2$As$_2$ \cite{PhysRevB.83.054514}, Ba(Fe$_{1-x}$Rh$_{x}$)$_2$As$_2$ \cite{PhysRevB.81.134512, PhysRevB.84.024509}, Ba(Fe$_{1-x}$Ni$_{x}$)$_2$As$_2$ \cite{PhysRevLett.108.247002}, Ba(Fe$_{1-x}$Co$_{x}$)$_2$As$_2$ \cite{PhysRevB.81.134512, PhysRevLett.104.057006, PhysRevLett.105.057001, PhysRevLett.103.087002, PhysRevLett.106.257001}

\subsection{Magneto-structural phase transition}
\begin{figure}[htbp]
	\centering
		\includegraphics[width=1.00\columnwidth]{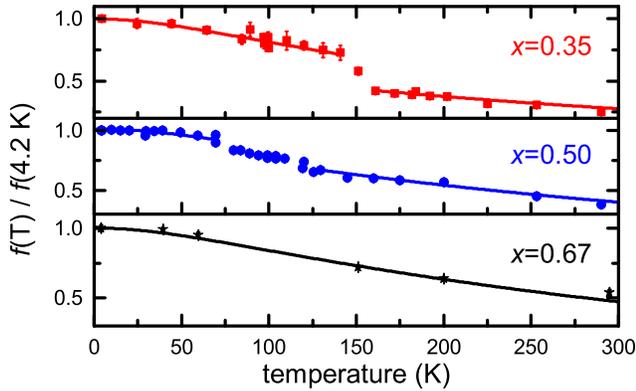}
	\caption{Temperature dependence of the relative recoilless fraction \textit{f}(\textit{T})/\textit{f}(4.2~K) including fits following Eq.~\ref{eq:debye}.}
	\label{fig:debye}
\end{figure}

The lattice dynamics of Ca$_{1-x}$Na$_x$Fe$_2$As$_2$ was investigated by analysing the temperature dependence of the M\"ossbauer-Lamb factor.
The recoilless fraction \textit{f} was extracted from the M\"ossbauer spectra using the absorption area method.\cite{Housley196429}
In the Debye-approximation,\cite{Barb} 

\begin{equation}
	f \propto \text{exp}\left \lbrace-{\frac{3E_R}{2k_B \theta_{D}} \left [ 1 + 4 \left (\frac{T}{\theta_{D}} \right )^2 \int_0^{\theta_{D} / T} \frac{x}{\text{e}^x -1}dx \right ]}\right \rbrace,
	\label{eq:debye}
\end{equation}
with Boltzmann constant \textit{k}$_B$, Debye temperature $\theta_{D}$ and the recoil energy $E_R$.
The temperature dependence of the relative recoilless fraction \textit{f}(\textit{T})/\textit{f}(4.2~K) is shown in Fig.~\ref{fig:debye}.
The data for temperatures above and below the phase transition were analyzed using Eq.~(\ref{eq:debye}) to investigate the influence of the magneto-structural transition on \textit{f}.
We accordingly define Debye temperatures $\theta_{D}^{\text{PM}}$ and $\theta_{D}^{\text{AFM}}$.
The step at 151~K for \textit{x}~=~0.35 and the gradual increase between 60~K and 125~K for \textit{x}~=~0.50 are attributed to an effective absorber-thickness effect.
 %due to a 
The magnetic phase transition leads to an increase of the absorption area compared to the paramagnetic region, due to a splitting of the resonance lines.\cite{Lang1963425, Barb}
The obtained Debye temperatures are shown in Tab.~\ref{tab:debye}.
\begin{table}[htbp]
	\begin{tabularx}{\columnwidth}{Y Y Y}\hline\hline
	\\
		\textit{x}		&	$\theta_{D}^{\text{PM}}$/K	&	$\theta_{D}^{\text{AFM}}$/K	 \\\\\hline \hline\\
	0.00	&	272\cite{0953-8984-23-25-255701}						&	271\cite{0953-8984-23-25-255701}				\\\hline
	0.35	&	200(32)				&	208(16)		\\\hline
	0.50	&	203(12)				& 219(28)	\\\hline
	0.67	&	213(6)				&  \\\\\hline\hline
	\end{tabularx}
	\caption{Debye temperatures $\theta_{D}$ obtained using Eq.~(\ref{eq:debye}) above (PM) and below (AFM) the magnetic phase transition.}
	\label{tab:debye}
\end{table}
As the Debye temperatures $\theta_{D}^{\text{PM}}$ and $\theta_{D}^{\text{AFM}}$ does not change significantly within error bars at the phase transition, we conclude that the lattice dynamics do not change upon the magneto-structural phase transition.
%Thus we define the Debye temperature of the sample $\theta_{D}$=~$\theta_{D}^{\text{PM}}$=~$\theta_{D}^{\text{AFM}}$.
%As the Debye temperatures $\theta_D$ in the paramagnetic and fully magnetically ordered state are of the same value within error bars, we conclude, that the lattice dynamics do not change upon the magneto-structural phase transition.
Also, the obtained Debye temperatures are constant within error bars for all \textit{x}.
Therefore, the lattice dynamics are independent of the Na-substitution level in the investigated substitution range 0.35~$\leq$~\textit{x}~$\leq$~0.67.
In comparison with the undoped compound CaFe$_2$As$_2$ with $\theta_{D}^{x=0}\approx$~270~K,\cite{0953-8984-23-25-255701} the lattice is softened considerably.

\begin{figure}[htbp]
	\centering
		\includegraphics[width=1.00\columnwidth]{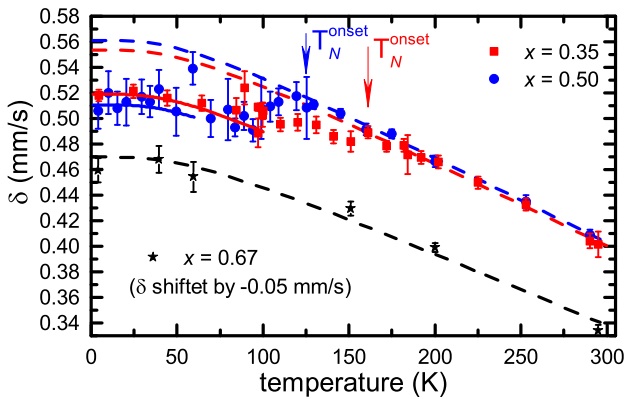}
	\caption{Temperature dependence of the isomer shift. The dashed (solid) lines are a fit in the paramagnetic (magnetically ordered) temperature regime using Eq.~(\ref{eq:cs}) with $\theta_{D}$~=~200~K, 203~K ,and 213~K for \textit{x}~=~0.35, 0.50, and 0.67, respectively, leading to $\delta^{\text{PM}}$(\textit{T}). The deviation from $\delta^{\text{PM}}$(\textit{T}) corresponds to the magneto-structural transition, which causes a change in the electron-density at the nucleus. Data and fit for \textit{x}~=~0.67 are shifted by $\delta~=~-0.05$~mm/s for clarity.}
	\label{fig:cs}
\end{figure}

To further study the magnetic properties at the magneto-structural phase transition, the temperature dependence of the isomer shift $\delta$ was analyzed.
$\delta$ is a measure for the electron density at the Fe-nucleus.
The temperature dependence of the isomer shift,\cite{Barb} which is shown in Fig.~\ref{fig:cs}, is a sum of the temperature-independent chemical shift $\delta_{\text{C}}$ and a temperature dependent contribution $\delta_{\text{R}}$(\textit{T}) due to the second-order Doppler shift
\begin{align}
	\delta (T) &= \delta_{\text{C}} + \delta_{\text{R}} (T),	\label{eq:cs} \\
	\delta_R (T) &= -\frac{9}{16} \frac{k_B}{M_{\text{eff}}~\mathrm{c}} \left [ \theta_{D} + 8T \left ( \frac{T}{\theta_{D}} \right )^3 \int_0^{\theta_{D} / T} \frac{x^3}{\text{e}^x -1}\text{d}x \right ],
\end{align}
where $M_{\text{eff}}$ denotes the effective mass of the $^{57}$Fe atom.
$\delta_{\text{C}}$ can be calculated using $\delta_{\text{C}} = \delta(0) - \delta_{\text{R}}(0)$.
To study the influence of the magneto-structural phase transition on the isomer shift, the temperature dependence of the isomer shift was analyzed in the following way:
$\delta$(\textit{T}) was analyzed in the paramagnetic state with a fixed $\theta_{D}$, obtained from Eq.~(\ref{eq:debye}).
Then, we extracted the temperature dependence from $\delta^{\text{PM}}$(\textit{T}).
In a third step we have checked whether $\delta^{\text{PM}}$(\textit{T}) can describe the temperature dependence of the isomer shift in the magnetically ordered state or whether systematic deviations from the behavior in the paramagnetic state occur.
%Therefore, both \textit{x}~=~0.35 and 0.50 were analyzed in the paramagnetic temperature regime using Eq.~(\ref{eq:cs}) with $\theta_{D}$=~200~K and $\theta_{D}$=~203~K.
$\delta^{\text{PM}}$(\textit{T}) reveals similar behavior for both samples with \textit{x}~=~0.35 and 0.50 leading to $\delta_{\text{C}}^{\text{PM}}$=~0.61(3)~mm/s and $\delta_{\text{C}}^{\text{PM}}$=~0.60(1)~mm/s.
%For simplicity reasons, only the resultant graph of x~=~0.35 is shown in Fig.~\ref{fig:cs}.
For \textit{x}~=~0.35, below the onset of the magnetic order, a deviation from the paramagnetic behavior is found.
$\delta_{\text{C}}^{\text{PM}}$ is reduced to a value of $\delta_{\text{C}}^{\text{AFM}}$=~0.57(1)~mm/s.
For \textit{x}~=~0.50, a reduction to a value of $\delta_{\text{C}}^{\text{AFM}}$=~0.56(1)~mm/s was observed.
Also, this reduction occurs over a wider temperature range as the magnetic transition is broader.
 %showing a strenghtening of the Fe-As intralayer coupling due to the magnetic component of the magneto-structural phase transition.
%This is supporting the result from DFT-calculations.\cite{PhysRevLett.102.037003}
These reductions in the chemical shifts correspond to an increase in the electron density at the nucleus.\cite{Barb}
The origin of this increase can be the structural or the magnetic phase transition.
The change from a tetragonal to an orthorhombic structure changes the lattice parameters and hence the volume of the unit cell, which may change the chemical shift.\cite{Barb}
An increase of the volume and hence an increase of the Fe-As distance results in an decrease of the electron density at the nucleus corresponding to an increased $\delta_{\text{C}}$.\cite{0953-8984-23-25-255701}
This behavior is for example observed in nonmagnetic FeSe, which shows an increase in $\delta_{\text{C}}$ of 0.006(1)~mm/s at the tetragonal-to-orthorhombic phase transition.\cite{Blachowski20101}
A magnetic phase transition may change the chemical shift, as observed in metallic iron at the Curie temperature, where no structural phase transition occurs.\cite{PhysRevLett.19.75}
The reduction of $\delta_{\text{C}}$ in metallic iron due to the magnetic phase transition is $\approx$~0.3~mm/s.\cite{PhysRevLett.19.75}
This may indicate that the origin of the change of the chemical shifts is of both magnetic and structural nature. 
%Therefore, the change of the chemical shifts are caused by the magneto-structural phase transition, where the structural changes result in an increase of $\delta_{\text{C}}$.\cite{0953-8984-23-25-255701, Blachowski20101}
%But as $\delta_{\text{C}}$ is reduced at the magneto-structural phase transition, the magnetic ordering has to reduce $\delta_{\text{C}}$.
However, previous M\"ossbauer measurements on the undoped compound show contradicting results with  either no change,\cite{0953-8984-23-14-145701} an increase\cite{0953-8984-23-25-255701}, or a decrease\cite{PhysRevB.83.134410} in the chemical shift.

For \textit{x}~=~0.67, the temperature dependence of $\delta$(\textit{T}) can be properly described by Eq.~(\ref{eq:cs}) and a value of $\delta_{\text{C}}$=~0.56(1)~mm/s was obtained.
Therefore, no signs of a magnetic or structural phase transition were found.

\subsection{Optimal doped Ca$_{0.33}$Na$_{0.67}$Fe$_2$As$_2$}

\begin{figure*}[htb]
	\centering
		\includegraphics[width=1.00\textwidth]{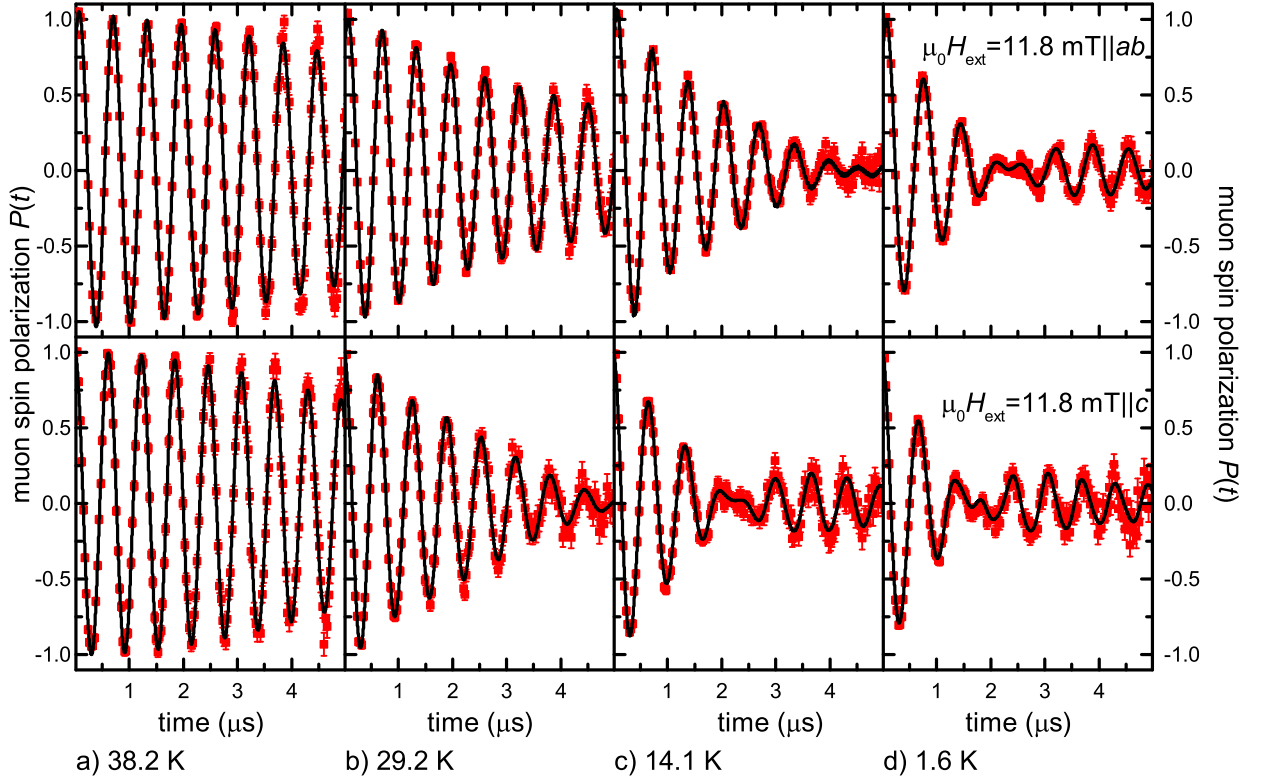}
	\caption{TF-$\mu$SR spectra for \textit{x}~=~0.67 with $\mu_0$\textit{H}$_{\text{ext}}$=11.8~mT$\|$\textit{ab} (upper row) and $\mu_0$\textit{H}$_{\text{ext}}$=11.8~mT$\|$\textit{c} (lower row) for temperatures above and below the superconducting transition temperature \textit{T}$_c~=$~34~K. The small (Gaussian) damping in spectrum (a) is attributed to the dipolar interaction of the muon spin with randomly distributed nuclear moments. The additional damping in spectra (b)--(d) is caused by the formation of the vortex lattice in the superconducting state and the associated internal magnetic field distribution \textit{n}(\textit{B}). It is clearly visible that the damping of the muon precession is stronger in the case $\mu_0$\textit{H}$_{\text{ext}}=~11.8~$mT$\|$\textit{c}.}
	\label{fig:SE3015_muSR_WEP+WEDL}
\end{figure*}

\begin{figure*}[htbp]
	\centering
	\captionsetup[subfloat]{position=top, parskip=0pt, aboveskip=0pt, justification=raggedright, labelseparator=none, farskip=0pt, nearskip=2pt, margin=0pt, captionskip=-15pt, font=normalsize, singlelinecheck=false}
	\subfloat[]{\includegraphics[width= 8.5cm]{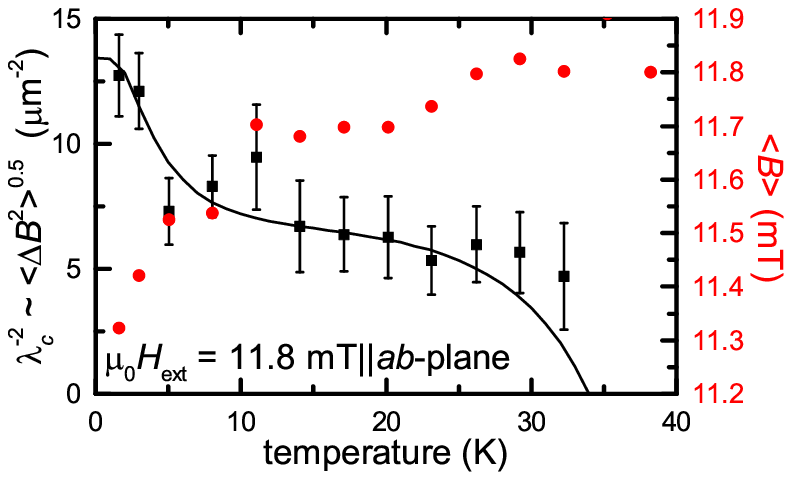} \label{fig:SE3015_WEP}}\quad
	\subfloat[]{\includegraphics[width= 8.5cm]{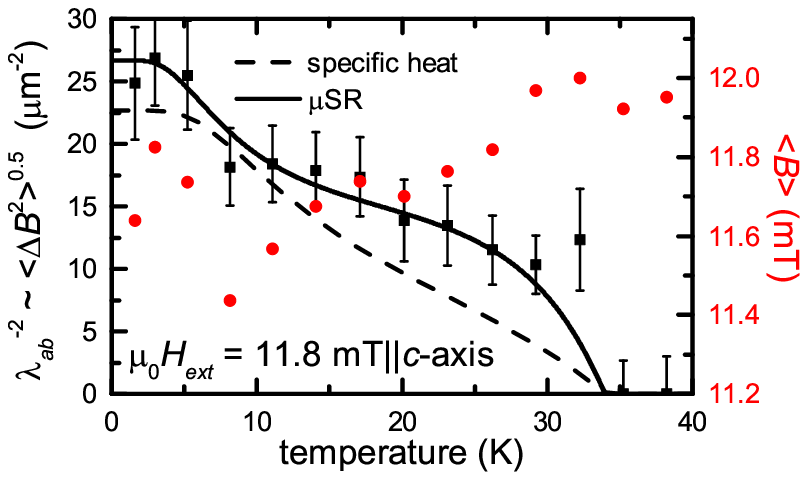} \label{fig:SE3015_WEDL}} 
	\caption{Temperature dependence of the magnetic penetration depth (a) $\lambda^{-2}_c (T)$ and (b) $\lambda^{-2}_{ab}(T)$ (proportional to the superfluid density) after field cooling in $\mu_0 H$~=~11.8~mT, including the fit (solid lines) with a phenomenological $\alpha$-model,\cite{Carrington2003205} and the average magnetic field determined by TF-$\mu$SR. The dashed curve in (b) displays the temperature dependence of $\lambda^{-2}$ with gap values of $\Delta_1$~=~2.35~meV, $\Delta_2$~=~7.5~meV, W($\Delta_1$)~=~0.75, and $\lambda$(0)~=~210~nm, obtained by specific-heat measurements\cite{PhysRevB.89.134507,PhysRevB.90.054524}. The reduction of the average magnetic field is caused by the diamagnetic shielding of the superconducting phase.}
	\label{fig:SE3015}
\end{figure*}
For Ca$_{0.33}$Na$_{0.67}$Fe$_2$As$_2$ (\textit{x}~=~0.67), susceptibility measurements evidence bulk superconductivity below \textit{T}$_c=$~34~K, as shown in Fig.~\ref{fig:susceptibility20Oe}.
M\"ossbauer spectra down to 5~K are shown in Fig.~\ref{fig:MBS_all}.
Above 151~K, a doublet structure with \textit{V}$_{zz}$=~10.3(2)~V/\AA$^2$ is observed.
This is consistent with a pure paramagnetic phase.
\textit{V}$_{zz}$ is constant within error bars down to lowest temperatures indicating the absence of a structural phase transition.
Below 60~K, a broadening of the spectra is observed.
This broadening indicates small magnetic fields, which were modeled using a Gaussian distribution with a first moment $\braket{B}$~=~0.
The standard deviation of this Gaussian distribution is constant within error bars, $\sigma(B)$~=~2.2(1)~T, above $T_c$ and decreases in the superconducting phase to 1.7(1)~T at 4.2~K.
This indicates a competition between magnetism and superconductivity. 
%consistent with the order parameter reduction in the coexistence regime for \textit{x}~=~0.50.
%Finite magnetic hyperfine fields are also seen in the non-magnetic superconductor LaOFeP ($T_c$~=~7~K), where a hyperfine field of 1.15(1)~T is observed.\cite{s63b1057}
Additionally, the spectra are nearly symmetric below 60~K and an angle $\theta$~=~55(5)$^{\circ}$ between the principal axis of the EFG and the magnetic hyperfine field is obtained, which is close to the magic angle.
ZF-$\mu$SR-experiments down to 5~K are shown in Fig.~\ref{fig:muSR_all_neu}.
The time evolution of the muon spin polarization exhibits a Gauss-Kubo-Toyabe depolarization above 75~K excluding any electronic magnetic order.
Below 40~K, a weak exponential relaxation supports short-range magnetic order in a small volume fraction with a MVF smaller than 20~\%.
By combining both local probes, the onset temperature of the weak magnetic order is estimated to be 60~K~$<$~$T_{N}^{\text{onset}}~<$~75~K.
However, room-temperature M\"ossbauer measurements as well as the sharp superconducting transition observed by magnetic-susceptibility measurements indicate a homogeneous sample.
This indicates that the weak magnetism is diluted and disordered and persistent even in the optimal doping regime, similar to other iron pnictides. \cite{1367-2630-11-5-055050, nmat2396, nmat2397}

For an investigation of the superconducting phase, TF-$\mu$SR measurements were performed in an external magnetic fields of $\mu_0 $\textit{H}$_{\text{ext}}$~=~11.8~mT perpendicular and parallel to the crystallographic \textit{c}--axis.
The magnetic field was applied at \textit{T}$>$\textit{T}$_c$ and the corresponding muon spin polarization is shown in Fig.~\ref{fig:SE3015_muSR_WEP+WEDL}(a).
The weak relaxation above \textit{T}$_c$ is caused by the dipole-dipole interaction of the muon spin with randomly distributed dense nuclear moments.
Additional damping is found in the case of a type-II superconductor for $\mu_0 $\textit{H}$_{c1} < \mu_0 $\textit{H}$_{\text{ext}} < \mu_0 $\textit{H}$_{c2}$ due to the vortex lattice formation.
The effect of the vortex lattice on the muon spin polarization is shown in Fig.~\ref{fig:SE3015_muSR_WEP+WEDL}(b)--(d).
The superconducting signal fraction is fully damped after a few $\mu$s and $\approx$~18~\% residual signal fraction is still oscillating with a precession frequency equal to the applied field at times $t$~$>$~3~$\mu$s.
Identifying this $\approx$18~\% signal fraction with the MVF obtained by ZF-$\mu$SR measurements, which is of equal value, shows that the internal magnetic fields are small compared to the 11.8~mT applied field.

The London penetration depth can be obtained by measuring the magnetic-field distribution within the vortex lattice and employing Eq.~(\ref{eq:brandt}).
Using the measurements with $\mu_0$\textit{H}$_{\text{ext}}$$\|$\textit{c}, the in-plane penetration depth $\lambda_{ab}$can be directly calculated.
%(0)=194(17)~nm 
For $\mu_0$\textit{H}$_{\text{ext}}$$\perp$\textit{c}, contributions from $\lambda_{ac}$ and $\lambda_{bc}$ are measured resulting in an effective magnetic penetration depth $\lambda_{c_{\text{eff}}}$.
Under the assumption of $\lambda_{a}~\approx~\lambda_{b}$, a value for $\lambda_{c}$ can be estimated using \cite{PhysRevLett.102.187005}
\begin{equation}
	\lambda_{ab}= \sqrt{\lambda_a \lambda_b} \approx \lambda_a \rightarrow \lambda_{c_{\text{eff}}} = \sqrt{\lambda_a \lambda_c} \rightarrow \lambda_c = \frac{\lambda_{c_{\text{eff}}}^2}{\lambda_a}.
\end{equation}
%where a value of $\lambda_{\text{c}}$(0)=280(46)~nm is obtained.
The resulting temperature dependence of the inverse squared London penetration depth, 1/$\lambda^2_{ab} (T)$ and 1/$\lambda^2_{c} (T)$, is shown in Fig.~\ref{fig:SE3015}(a) and Fig.~\ref{fig:SE3015}(b), respectively, together with the average internal field $\braket{B}$, which shows a reduction due to the diamagnetic shielding below the superconducting transition.
1/$\lambda^2$(T) was modeled using  the phenomenological $\alpha$-model including two independent superconducting gaps with \textit{s}-wave symmetry and a fixed $T_c~=~34$~K.\cite{Carrington2003205}
The results are shown in Tab.~\ref{tab:gap} as well as results of ARPES\cite{PhysRevB.87.094501} (\textit{x}~=~0.67) and specific-heat measurements\cite{PhysRevB.89.134507} (\textit{x}~=~0.68) on single crystals.
\begin{table}[htpb]
	\begin{tabular}{ccccc}\hline\hline\\
		&	$\Delta_1$(0)/meV&	$\Delta_2$(0)/meV&	\textit{W}($\Delta_1$)	&	$\lambda$(0)/nm \\\\\hline\hline\\
		$\lambda_c$	&	0.57(8)	&	6.7(1.3)	&	0.49(4)	&	280(46)\\\hline
		$\lambda_{ab}$	&	0.8(3)	&	6(1)	&	0.46(8)	&	194(17) \\\hline
		ARPES\cite{PhysRevB.87.094501}		&	2.3 & 7.8 &&\\\hline
		spec. heat\cite{PhysRevB.89.134507,PhysRevB.90.054524}	&	2.35 & 7.5	& 0.75\cite{PhysRevB.89.134507} & 210(10)\cite{PhysRevB.90.054524}\\\\\hline\hline
	\end{tabular}
	\caption{Values of the superconducting gap and the magnetic penetration depth obtained by a phenomenological $\alpha$-model analysis of 1/$\lambda^2_c(T)$ and 1/$\lambda^2_{ab}(T)$. $\Delta_i$(0) denotes the zero temperature values of the gaps. \textit{W}($\Delta_1$) and  \textit{W}($\Delta_2$)=1$-$\textit{W}($\Delta_1$) are the corresponding weighting factors. $\lambda$(0) denotes the zero-temperature penetration depth.}
	\label{tab:gap}
\end{table}
%The obtained values for the superconducting energy gaps are in good agreement with recent ARPES\cite{PhysRevB.87.094501} and specific-heat measurements\cite{PhysRevB.89.134507} on single crystals from the same batch with the same composition.
Disorder in the vortex lattice would artificially reduce the magnetic penetration depth due to the broadening of \textit{n}(\textit{B}).
Therefore, $\lambda_{ab}$ and $\lambda_{c}$ strictly describe lower limits (and $\lambda^{-2}$ an upper limit).
Therefore, the obtained values for the magnetic penetration depths are reduced compared to the values obtained by, e.g., specific-heat measurements.
To illustrate this effect, the temperature dependence of $\lambda^{-2}_{\text{spec. heat}}(T)$ with the corresponding parameter ($\Delta_1$~=~2.35~meV, $\Delta_2$~=~7.5 meV, W($\Delta_1$)~=~0.75 and $\lambda$(0)~=~210~nm) obtained by specific-heat measurements\cite{PhysRevB.89.134507, PhysRevB.90.054524} is plotted in Fig.~\ref{fig:SE3015_WEDL}.
%This is illustrated in Fig.~\ref{fig:SE3015_WEDL}, where, in addition to the by $\mu$SR-experiments obtained temperature dependence of 1/$\lambda^2$, the corresponding graph with parameter obtained by specific-heat measurements\cite{PhysRevB.89.134507, PhysRevB.90.054524} is plotted.
It is clearly visible that $\lambda^{-2}_{\text{spec. heat}}(T)~<~\lambda^{-2}_{\mu\text{SR}}(T)$ for $T~<~T_c$.
This underestimation of the magnetic penetration depth in the $\mu$SR experiments may result in a different temperature dependence of $\lambda^{-2}(T)$ and therefore in different gap sizes and weighting factors.
Additionally, Johnston \textit{et al.} considered the interband coupling of the superconducting bands and found an intermediate coupling strength.\cite{PhysRevB.89.134507}
The $\alpha$-model used in this work considers two noninteracting superconducting bands, which may also explain the different parameter values.
Nevertheless, taking into account the limitation of magnetic penetration depth measurements by means of $\mu$SR experiments and of the used $\alpha$-model, the obtained parameter for $\Delta_1(0)$, $\Delta_2(0)$, $W(\Delta_1)$, and $\lambda(0)$ are in good agreement with the values obtained by ARPES and specific heat experiments.\cite{PhysRevB.87.094501, PhysRevB.89.134507, PhysRevB.90.054524}

The anisotropy of the magnetic penetration can be calculated by under the assumption that $\lambda_{a}~\approx~\lambda_{b}$ by\cite{PhysRevLett.102.187005}
\begin{equation}
	\gamma_{\lambda} = \frac{\lambda_c}{\lambda_{ab}}.
\end{equation}
%The temperature dependence of $\gamma_{\lambda}$ is shown in Fig.~\ref{fig:SE3015_anisotropy}.
A temperature-independent value of $\gamma_{\lambda}$~=~1.5(4) is the smallest observed among the 122 pnictides indicating a more 3D-behavior.\cite{PhysRevB.89.134507}
This behavior is consistent with the temperature-independent value of $\gamma$~=~1.85(5) for the anisotropy of the upper critical fields.\cite{PhysRevB.84.064533}

\section{Summary and conclusions}
\begin{figure}[htbp]
	\centering
	\includegraphics[width=1.00\columnwidth]{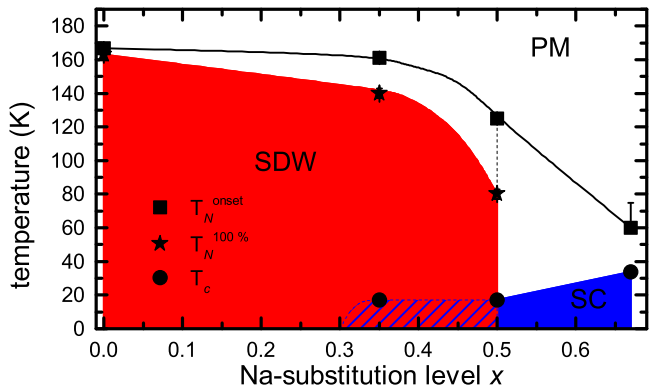}
	\caption{Updated phase diagram of Ca$_{1-x}$Na$_x$Fe$_2$As$_2$. The magnetic order is suppressed as a function of the Na-substitution level \textit{x}. The temperature width of the magnetic phase transition $\Delta$\textit{T}~=~\textit{T}$_N^{\text{onset}}$ $-$ \textit{T}$_N^{100~\%}$ increases as a function of \textit{x} due to the enhanced degree of disorder due to the Na substitution. For \textit{x}~=~0.35, 0.50, and 0.67 nanoscopic coexistence of magnetic order and superconductivity is found below $T_c$.}
	\label{fig:PDNeu}
\end{figure}

In summary, we performed muon spin relaxation and M\"ossbauer experiments on Ca$_{1-x}$Na$_x$Fe$_2$As$_2$ single crystals with \textit{x}~=~0.00, 0.35, 0.50, and 0.67 resulting in an updated phase diagram, which is shown in Fig.~\ref{fig:PDNeu}.
The substitution of Ca by Na reduces the onset of the magnetic ordering from \textit{T}$_N^{\text{onset}}$~=~167(2)~K to 161(2)~K and 125(3)~K while the magnetic phase transition temperature width $\Delta$\textit{T}~=~\textit{T}$_N^{\text{onset}}$ $-$ \textit{T}$_N^{100~\%}$ increases from 4~K to 21~K and 45~K for \textit{x}~=~0.00, 0.35, and 0.50, respectively.
The muon spin precession frequency $\nu$ as well as the magnetic hyperfine field \textit{B}$_{\text{hf}}$, which are proportional to the magnetic order parameter, are reduced as a function of the Na-substitution level.
Both $\mu$SR as well as M\"ossbauer spectroscopy indicate an increased tilting of the magnetic structure upon doping.
The magnetic phase transition is accompanied by a structural phase transition.
The lattice dynamics does not change at the magneto-structural phase transition.
%By analysing the lattice dynamics, we found no change upon the magneto-structural phase transition.
%For higher Na-concentrations, magnetic order is completely suppressed.
Magnetic susceptibility measurements indicate superconductivity in parts of the sample volume for \textit{x}~=~0.35, whereas the sample with \textit{x}~=~0.50 shows bulk superconductivity.
Therefore, as 100\% of the sample is magnetically ordered, coexistence of magnetic order and superconductivity in parts (\textit{x}~=~0.35) or in the whole sample (\textit{x}~=~0.50) was observed.
A strong reduction of the magnetic order parameter, as found in the Ba$_{1-x}$Na$_x$Fe$_2$As$_2$ series,\cite{Maeter} is not observed for Ca$_{1-x}$Na$_{x}$Fe$_2$As$_2$ with \textit{x}~=~0.35.
For \textit{x}~=~0.50, a small reduction of $\approx$~7~\% was observed.
We applied a Landau theory to describe the reduction of the magnetic order parameter showing that the magnitude of the reduction depends on the coupling strength and the $T_c/T_N$ ratio.
A linear relation between the reduction of $|\vec{M}_{\text{co}}|^2/|\vec{M}_0|^2(T=0)$ on $T_c/T_N$ has been found for several superconducting 122-pnictide systems that microscopic coexistence.

For \textit{x}~=~0.67, diluted and weak magnetism below 60--75~K as well as bulk superconductivity with $T_c~=$~34~K is found.
The \textit{s}-wave symmetry of the two superconducting gaps as well as the value of the larger gap agrees well with recent ARPES-and specific-heat measurements.\cite{PhysRevB.87.094501,1402.5875}
%Further investigations at Na-substitution levels between 0.50 and 0.67 are usefull to determine, at which \textit{x} the long-range magnetic order is fully suppressed and the diluted magnetism appears.
%Additionally, the origin of the magnetic field at the Fe-nuclei for \textit{x}~=~0.67 is still unclear.
%In conclusion, Ca$_{1-x}$Na$_x$Fe$_2$As$_2$ shows magnetic order up to highest Na-substitution levels.
%For a deeper understanding of the \textit{x}-dependency of the suppression of the magnetic order as well as the interaction of the magnetic and superconducting order parameter, more stoichiometries are needed.
\acknowledgments
This work was funded by the German Science Foundation (DFG) within the research training group GRK 1621 as well as under the Projects WU595/3-1 (S. W.), BE1749/13 and BU887/15-1 (B. B.), and SA 2426/1-1 (R. S.).
%R.S. is grateful to the DFG for financial support with the grant number SA 2426/1-1.
Part of this work was performed at the Swiss Muon Source at the Paul Scherrer Institute, Switzerland.

\bibliography{CaNaFe2As2}% Produces the bibliography via BibTeX.

\end{document}